\documentclass[
    ,final            
  ]
  {aipproc}

\layoutstyle{6x9}
\usepackage{times}
\usepackage{bm}
\newcommand{\kA}{k^{\rm A}}
\newcommand{\kc}{k^{\rm c}}
\newcommand{\EQ}{\begin{equation}}
\newcommand{\EN}{\end{equation}}
\newcommand{\EQA}{\begin{eqnarray}}
\newcommand{\ENA}{\end{eqnarray}}
\newcommand{\eq}[1]{(\ref{#1})}

\newcommand{\Eq}[1]{equation~(\ref{#1})}

\newcommand{\Eqss}[2]{equations~(\ref{#1})--(\ref{#2})}

\newcommand{\Fig}[1]{Fig.~\ref{#1}}
\newcommand{\FFig}[1]{Figure~\ref{#1}}
\newcommand{\Tab}[1]{Table~\ref{#1}}
\newcommand{\Figs}[2]{Figs~\ref{#1} and \ref{#2}}

\newcommand{\bra}[1]{\langle #1\rangle}

{}

\newcommand{\meanBB}{\overline{\bf{B}}}

\newcommand{\meanUU}{\overline{\bf{U}}}
\newcommand{\nab}{{\bm{\nabla}}}
{}
{}
{}
{}
{}
{}
{}

\newcommand{\yyy}{\hat{\bm y}}

\newcommand{\qq}{{\bm{q}}}

\newcommand{\uu}{{\bm{u}}}
\newcommand{\BB}{{\bm{B}}}

\newcommand{\JJ}{{\bm{J}}}

\newcommand{\AAA}{{\bm{A}}}
\newcommand{\aaaa}{{\bm{a}}}
\newcommand{\bb}{{\bm{b}}}

\newcommand{\ff}{\mbox{\boldmath $f$} {}}

\newcommand{\FF}{{\bm{F}}}

\newcommand{\LLLL}{{\sf{\bm{L}}}}

\newcommand{\ii}{{\rm i}}

\newcommand{\DDD}{{\cal D} {}}
\newcommand{\dd}{{\rm d} {}}

\def\Pm{\mbox{\rm Pr}_M}

\def\threehalf{{\textstyle{3\over2}}}

\newcommand{\G}{\,{\rm G}}

\newcommand{\kG}{\,{\rm kG}}

\newcommand{\s}{\,{\rm s}}

\newcommand{\m}{\,{\rm m}}

\newcommand{\kg}{\,{\rm kg}}

\begin{document}

\title{Shearing and embedding box simulations of the magnetorotational
instability}

\author{Axel Brandenburg}{
  address={Nordita, Blegdamsvej 17, DK-2100 Copenhagen \O, Denmark}
}

\author{Boris Dintrans}{
  address={Obs. Midi-Pyr\'en\'ees, CNRS UMR5572, 14 avenue
Edouard Belin, 31400 Toulouse, France}
}

\author{Nils Erland L. Haugen}{
  address={DAMTP, University of Cambridge,
Wilberforce Road, Cambridge CB3 0WA, UK, and\\
Department of Physics, The Norwegian University of Science
and Technology, H{\o}yskoleringen 5, N-7034 Trondheim, Norway
}
}

\begin{abstract}
Two different computational approaches to the magnetorotational instability
(MRI) are pursued: the shearing box approach which is suited for local
simulations and the embedding box approach whereby a Taylor Couette flow
is embedded in a box so that numerical problems with the coordinate singularity
are avoided. New shearing box simulations are presented and differences
between regular and hyperviscosity are discussed. Preliminary simulations
of spherical nonlinear Taylor Couette flow in an embedding box are presented
and the effects of an axial field on the background flow are studied.
\end{abstract}

\maketitle

\section{Introduction}

It is now generally accepted that the magnetorotational instability
(MRI) is the main agent driving turbulence in accretion discs
\cite{BAL98}.
This instability provides a good explanation for the turbulence in
accretion discs which might otherwise be hydrodynamically stable
\cite{BAL96}.
The historical developments of the understanding of this instability
have appropriately been reviewed elsewhere in this book.
Here we focus mainly on results within the shearing sheet
approximation, which is the relevant extension of a periodic geometry
to a shearing environment.
Like with periodic ``boundary'' conditions, there is actually no boundary,
and every point in the domain is equivalent to any other point.
This avoids boundary layers which
is extremely useful for simulations, but it is also astrophysically
more relevant because we do not want explicit boundaries in a local
domain that is supposed to represent a subvolume within the global disc.
In the following we present a discussion of the nonaxisymmetric MRI in
the shearing sheet approximation using the Rayleigh quotient to define
in a convenient way an ``instantaneous'' growth rate of otherwise only
transient growth.
In this paper we also calculate the nonlinear evolution of the MRI,
reviewing both old simulations using subgrid scale modeling and new
direct simulations where the ordinary viscosity and diffusion
operators are used with uniform coefficients.
Finally we turn to the issue of spherical Couette flow and present
preliminary simulations in order to address recent experimental
results suggestive of MRI in liquid sodium.

\section{Axisymmetric vs nonaxisymmetric MRI}

In order to obtain the dispersion relation for the MRI with a vertical
magnetic field it suffices to consider the linearized pressureless
one-dimensional
momentum equation and the linearized induction equation for the two
components in the rotational ($x,y$) plane, i.e.\
\EQ
\dot{u}_x-2\Omega u_y'=v_{\rm A} b_x',
\EN
\EQ
\dot{u}_y+(2-q)\Omega u_x'=v_{\rm A} b_y',
\EN
\EQ
\dot{b}_x=v_{\rm A} b_x',
\EN
\EQ
\dot{b}_y+q\Omega b_x'=v_{\rm A} b_y',
\EN
where dots and primes denote derivatives with respect to $t$ and $z$,
respectively, $u$ and $b$ are linearized velocity and magnetic field,
and $q$ denotes the steepness of the rotation law with $\Omega\sim R^{-q}$
where $q=3/2$ for purely keplerian discs.
In our local cartesian model, $y$ is the streamwise direction and $x$ is
the cross-stream direction.

One could have retained the $u_z$ component of the momentum equation
where the pressure gradient would enter, together with the continuity
equation, but these two equations decouple from those considered already.
The resulting new modes are the fast magnetosonic waves, which are
of no particular interest in the present context.

\begin{figure}
  \label{FBalbusHawley}
  \includegraphics[height=.4\textheight]{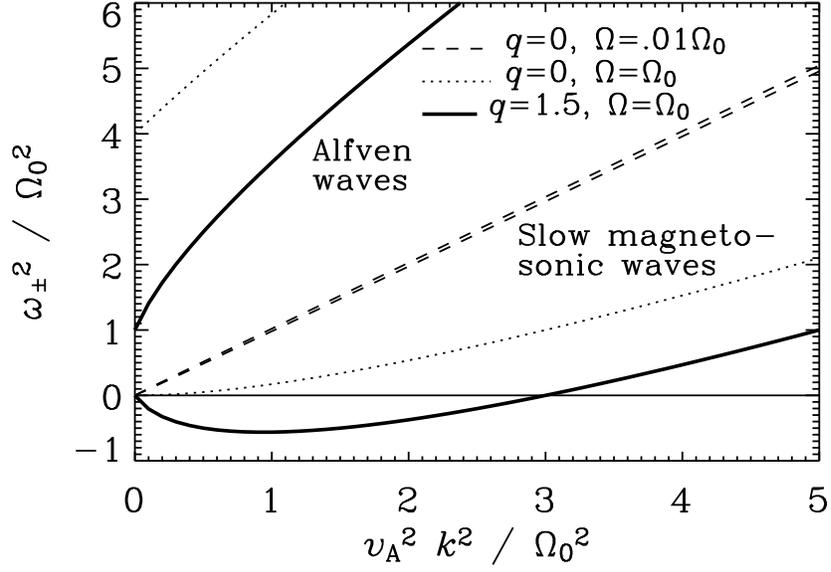}
  \caption{Solutions of the dispersion relation \eq{DisperSol} for accretion
disc parameters ($q=3/2$, thick solid lines), compared with the dispersion
relation for a rigidly rotating disc ($q=0$, dotted lines) and an almost
nonrotating disc ($q=0$ and $\Omega=0.01\Omega_0$, dashed lines).
Here, $\Omega_0^2=GM/R^3$ is the keplerian value at radius $R$.
The upper set of curves denotes Alfv\'en waves and the lower set slow
magnetosonic waves.
Note that in the completely nonrotating case ($\Omega=0$) the Alfv\'en
and slow magnetosonic waves are degenerate.
}
\end{figure}

Assuming that the solution is proportional to
$e^{\ii kz-\ii\omega t}$, we obtain the dispersion relation
\cite{BAL91,BAL98} in the form
\EQ
\omega^4-2\omega^2\left[v_A^2k^2+(2-q)\Omega^2\right]
+v_A^2k^2\left(v_A^2k^2-2q\Omega^2\right)=0.
\label{disper_BH}
\EN
This is a bi-quadratic equation with altogether 4 solutions, corresponding
to two different branches where $\omega$ can have either sign on each
of them.
The two branches are
\EQ
\omega_\pm^2=v_A^2k^2+(2-q)\Omega^2
\pm\Omega\sqrt{4v_A^2k^2+(2-q)^2\Omega^2}.
\label{DisperSol}
\EN
The upper branch corresponds to Alfv\'en waves and the lower branch
corresponds to slow magnetosonic waves.
The fast magnetosonic waves have been eliminated by using the pressureless
momentum equation.
In the range $0<v_{\rm A}^2k^2<3$ the slow magnetosonic waves can become
unstable, i.e.\ $\omega_-^2<0$, corresponding to an
exponentially growing solution
with growth rate $\mbox{Im}\,\omega_-$.
The maximum growth rate is reached when $v_A^2k^2={15\over16}\Omega^2$,
and the corresponding value of $\omega_-^2$ is then
$\mbox{Im}\,\omega_-={3\over4}\Omega$; see \Fig{FBalbusHawley}.

In the nonaxisymmetric case the situation is not quite as simple,
because now $\partial/\partial y\neq0$, and this leads to the
occurrence of a new term with a non-constant coefficient.
This comes from the $\meanUU\cdot\nab$ advection terms in all
four equations, because $\meanUU=Sx\yyy$.
There are two different ways of dealing with this problem.
One possibility is to abandon the assumption of harmonic
solutions in the $x$ direction, i.e.\ one assumes
$\uu=\hat{\uu}(x)e^{\ii k_yy+\ii k_zz-\ii\omega t}$.
This approach has been taken by \cite{OGI96}, for example.
Another possibility is to assume shearing-periodic boundary conditions.
The way then to get rid of the $x$ dependence in the $Sx\partial/\partial y$
term is by assuming $k_x$ to depend on time, because then the
$\partial/\partial t$ term pulls down a factor $\dot{k}_xx$ that can
be arranged such as to cancel the $Sx k_y$ term that results from
the $Sx\partial/\partial y$ term.
(Here, as before, the dots denote a derivative with respect to $t$.)
This requires $\dot{k}_xx+Sxk_y=0$, i.e.\
\EQ
k_x(t)=k_{x0}-Stk_y,
\EN
where $k_{x0}$ is some initial values of $k_x$.
Using the ansatz
\EQ
\qq(x,y,z,t)=A\,\mbox{Re}\,\hat{\qq}(t)\,
\exp\left[\ii k_x(t)x+\ii k_yy+\ii k_zz\right],
\label{expansion}
\EN
where the hats denote the shearing sheet expansion,
$A$ is an amplitude factor, and
$\hat{\qq}=(\hat{u}_x,\hat{u}_y,\hat{u}_z,\hat{b}_x,
\hat{b}_y,\hat{b}_z,\hat{\Lambda})^T$
is the state vector with $\Lambda=c_{\rm s}\ln\rho$,
$\bb=\BB/\sqrt{\mu_0\rho_0}$,
the partial differential equation
\EQ
\ii{\partial\qq\over\partial t} 
+\ii Sx{\partial\qq\over\partial y}
=\LLLL\qq,
\label{q_eqn_with_S}
\EN
with non-constant coefficients, turns into an ordinary differential equation
with constant coefficients
\cite{GOL65,BAL92a},
\EQ
\ii{\dd\hat{\qq}\over\dd t}=\hat{\LLLL}\hat{\qq}.
\label{ode}
\EN
Without going into the details of the derivation we simply state the
governing matrix, $\hat{\LLLL}$, for the case with uniform density
$\rho_0$, a uniform magnetic field in the $y$ direction, $\yyy B_0$,
and Alfv\'en speed $v_{\rm A}=B_0/\sqrt{\mu_0\rho_0}$ and an isothermal
equation of state,
\EQ
\hat{\LLLL}=\pmatrix{
0&2\ii\Omega&0&-\kA_y&\kA_x&0&\kc_x\cr
-2\ii\Omega^{\rm S}&0&0&0&0&0&\kc_y\cr
0&0&0&0&\kA_z&-\kA_y&\kc_z\cr
-\kA_y&0&0&0&0&0&0\cr
\kA_x&0&\kA_z&\ii S&0&0&0\cr
0&0&-\kA_y&0&0&0&0\cr
\kc_x&\kc_y&\kc_z&0&0&0&0}\!,
\EN
where we have used the abbreviations $2\Omega^{\rm S}=2\Omega+S$,
$\kA_i=k_i v_{\rm A}$, $\kc_i=k_i c_{\rm s}$.
In the stratified case with a uniform background field, the system
of governing equations has variable coefficients. Therefore we
only consider here the case of constant density.
Note that $\hat{\LLLL}$ is hermitian if $S=0$.
This property ensures that all eigenvalue are stable in that case.

\begin{figure}
\includegraphics[height=0.34\textheight]{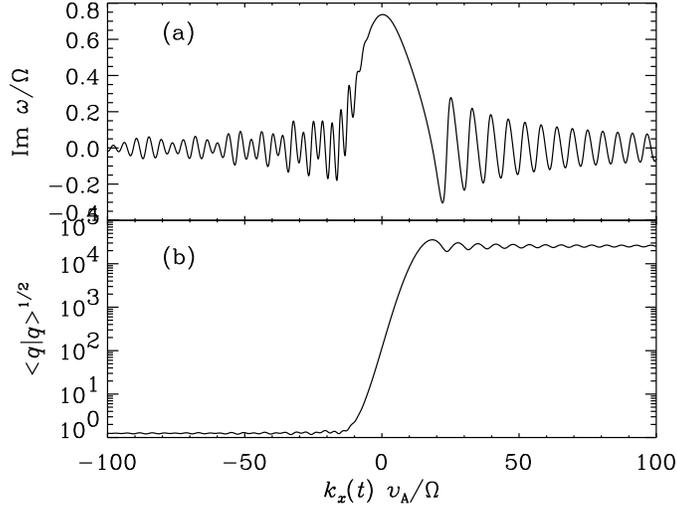}
\caption{Evolution of the imaginary part of $\omega$ (a) and the norm of
$\qq$ (b) for
$v_{\rm A}k_y/\Omega=1$.
Note that transient amplification is only possible during the time
interval when $|v_{\rm A}k_x(t)|/\Omega10$.}
\label{Fpres_typ}
\end{figure}

The set of equations \eq{ode} were already investigated
by \cite{BAL92a} using numerical integration.
They looked at the evolution $\hat{\qq}$ and found a transient behavior
that depends on initial conditions.
\cite{FOG94,FOG95} studied the nonaxisymmetric stability
in the presence of density stratification which also gives rise to
the Parker instability.
A convenient method to calculate the growth rates of the MRI is
in terms of the Rayleigh quotient (see \cite{BRD01})
\EQ
\omega(t)={\bra{\hat{\qq}|\hat{\LLLL}\hat{\qq}}\over\bra{\hat{\qq}|\hat{\qq}}},
\label{defomegat}
\EN
where $\displaystyle \bra{\aaaa|\bb}=\sum_{i=1}^N a^*_ib_i$ defines
a scalar product.
To ensure solenoidality of the magnetic
field, we calculate $\hat{b}_x$ for the initial perturbation from
\cite{BAL92a}
\EQ
\hat{b}_{x}(0)=-(k_y\hat{b}_y+k_z\hat{b}_z)/k_{x0}.
\EN
The results for the keplerian case, $S/\Omega=-3/2$, are shown in
\Figs{Fpres_typ}{Fppres}. In agreement with earlier work
\cite{BAL92a,KIM00}, the maximum growth rate
agrees with the Oort $A$-value \cite{BAL92b}, which is
$-S/2$, or ${3\over4}\Omega$ for keplerian rotation; see also
\Fig{FBalbusHawley}.

\begin{figure}[t!]
\includegraphics[height=0.28\textheight]{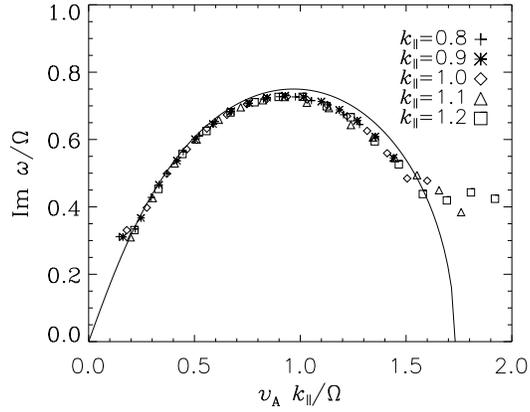}
\caption{Growth rates as a function of Alfv\'en speed, in units
of $\Omega/k_\parallel$ where $k_\parallel$ denotes $k_y$ in the
nonaxisymmetric case and $k_z$ in the axisymmetric one. Symbols mark the
results obtained from the Rayleigh quotient method
for the nonaxisymmetric instability with $k_y = [0.8,1.2]$.
For $v_{\rm A} k_y/\Omega > 1.7$ our technique fails
to yield reliable values of $\omega$ and the noisy oscillations
seen in \Fig{Fpres_typ} become dominant. The solid lines indicates the 
result for the axisymmetric MRI (Eq. \ref{DisperSol}).}
\label{Fppres}
\end{figure}

\begin{figure}[t!]
\includegraphics[height=0.40\textheight]{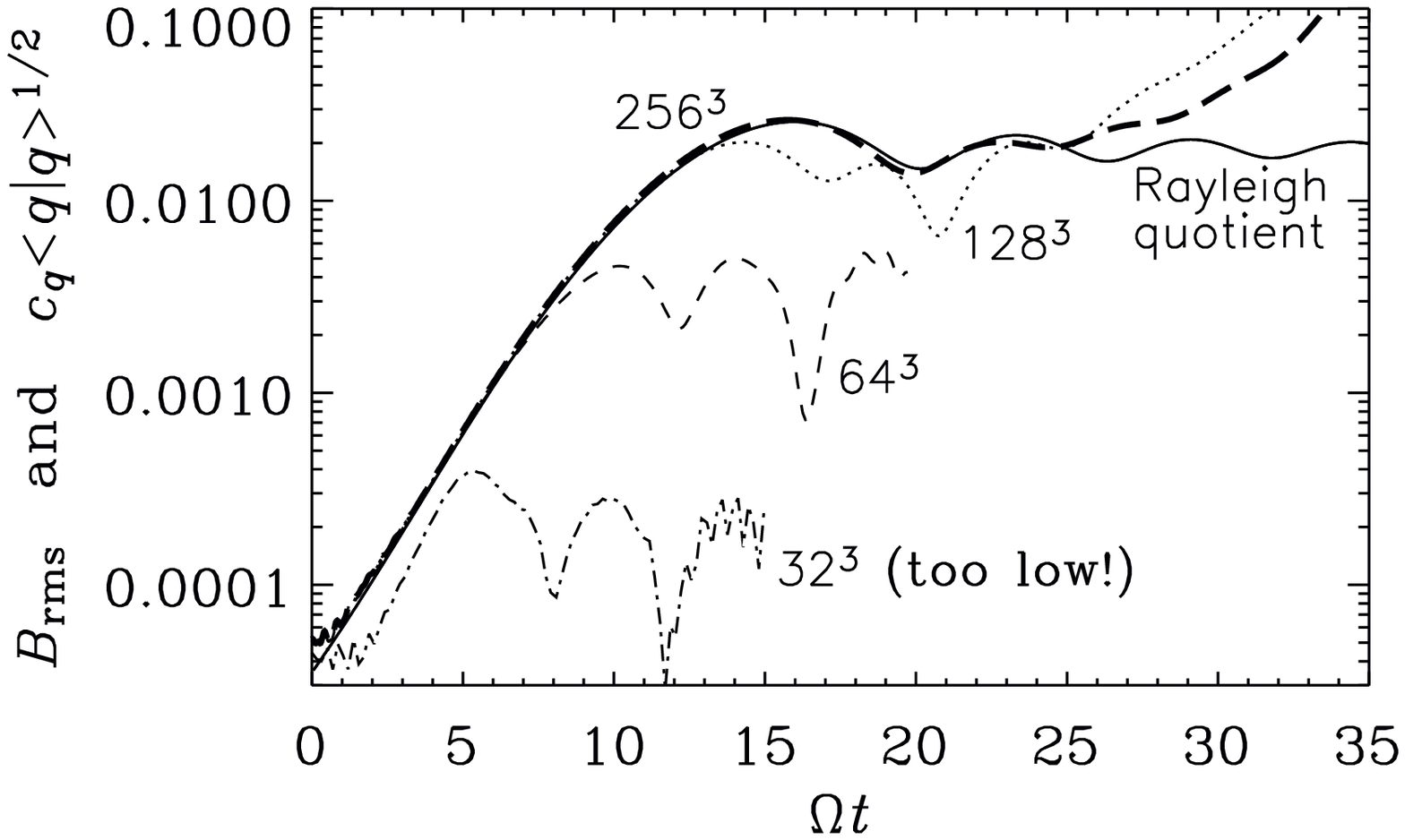}
\caption{Transient amplification of the magnetic field by the
nonaxisymmetric magnetorotational instability.
The solid line shows the result from the Rayleigh quotient method
while the broken lines give the result from direct three-dimensional
simulations with zero viscosity and zero resistivity.
The square root of the Rayleigh quotient has been scaled by a factor
$c_q=3.5\times10^{-5}$ to make it overlap with $B_{\rm rms}$ curve.
A resolution of only $32^3$ meshpoints is completely insufficient to
resolve even the beginning of the instability.
At least $256^3$ meshpoints are required to resolve the maximum.
After $\Omega t>17$ even the simulation with $256^3$ meshpoints
becomes under-resolved.}
\label{pcomp_col}
\end{figure}

Thus, we see that in the nonaxisymmetric case the MRI shows great
similarity with the axisymmetric counterpart.
However, in the present case of the shearing sheet approximation it
is strictly speaking only a transient.
This becomes more obvious when looking at the evolution of
$\bra{\qq|\qq}^{1/2}$; see \Fig{pcomp_col} where we see an increase
over about three orders of magnitude.
In this figure we also show the resulting evolution of the
root-mean-square magnetic field from a direct simulation
of the shearing sheet equations, where we also adopt an isothermal
gas with constant sound speed.
The continuity equation is written in terms of the logarithmic density,
$\Lambda\equiv\ln\rho$,
\EQ
{\DDD\Lambda\over\DDD t}=-\uu\cdot\nab\Lambda-\nab\cdot\uu,
\label{dlnrhodt}
\EN
and the induction equation is solved in terms of the magnetic vector
potential $\AAA$, where $\BB=\nab\times\AAA$, and
\EQ
{\DDD\AAA\over\DDD t}=\uu\times\BB+\threehalf\Omega_0 A_y \hat{x}
+\eta\nabla^2\AAA,
\label{dAdt}
\EN
$\eta$ being the magnetic diffusivity and $\Omega_0$ being the background 
rotation at the reference radius
(the derivation of the shear term in this form is given in
\cite{BRA95}).
The momentum equation is solved in the form
\EQ
{\DDD\uu\over\DDD t}=-\uu\cdot\nab\uu
-c_{\rm s}^2\nab\Lambda+{\JJ\times\BB\over\rho}
+\FF_{\rm visc}+\ff,
\label{dudt}
\EN
where $\DDD/\DDD t=\partial/\partial t+u_y^{(0)}\frac{\partial}{\partial y}$ 
 and $u_y^{(0)}=-(3/2)\Omega_0 x$ is the velocity in the y-direction due to 
the shear flow.
Furthermore $\JJ=\nab\times\BB/\mu_0$ is the current density,
$\BB$ is the magnetic field, $\mu_0$
is the vacuum permeability and
$\FF_{\rm visc}$ is the viscous force.

\begin{figure}[t]
\includegraphics[height=.2\textheight]{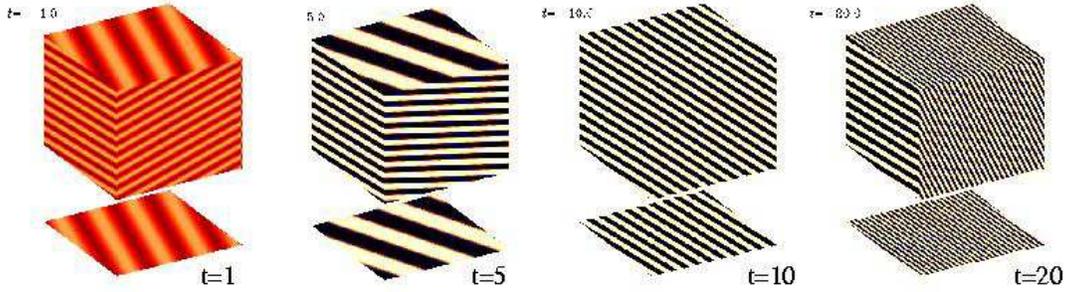}
\caption{
Images of the vertical component of the magnetic field, $B_z$, for
different values of $t$, where the field at $t=0$ corresponds to that
given by \Eq{initcond}.
}\label{img}\end{figure}

The initial condition for the 3-dimensional direct simulation is obtained
by evolving linearized shearing sheet equations for $k_y=1$ and $k_z=10$
to the point where $k_x(t_0)=-5$. [The size of the domain is $(2\pi)^3$.]
For definitiveness, we reproduce here the numerical values in
\Eq{expansion}:
\EQ
\hat{\uu}=\pmatrix{
-0.3108-0.0366\ii\cr
-0.4610-0.0542\ii\cr
-0.0883-0.0104\ii},\quad
\hat{\bb}=\pmatrix{
+0.0683-0.5813\ii\cr
-0.0393+0.3340\ii\cr
+0.0347-0.2950\ii},
\label{initcond}
\EN
and the logarithmic density is given by $\hat{\Lambda}=0.042-0.3647\ii$.
The amplitude is chosen to be $A=10^{-4}$.
\FFig{img} shows images of $B_z$ on the periphery of the simulation domain.
The simulations have been carried out using the {\sc Pencil Code}\footnote{
\url{http://www.nordita.dk/software/pencil-code}} which is a high-order
finite-difference code (sixth order in space and third
order in time) for solving the compressible hydromagnetic equations.

The way how this transient amplification can lead to sustained
growth is through mode coupling, which is not considered in the present
analysis.
Relevant mode couplings could come about either through nonuniformities
in the cross-stream or $x$ direction and boundary conditions, or through
nonlinearities.
In the shearing sheet approximation only the latter seems a viable
possibility, and this is probably the mechanism through which the early
shearing sheet simulations (e.g.\ \cite{HAW95,MAT95,BRA95})
produced sustained
turbulence.

\section{Local disc simulations and dynamos}

The magneto-rotational instability is believed to be of great importance
in connection with accretion discs.
Here, gas spirals gradually into the center.
This is possible mainly because of magnetic stresses, in particular those
resulting from the small scale fields, $\overline{b_rb_\phi}$.
They tap potential energy which gets converted into kinetic and magnetic
energies, and eventually into heat which gets radiated away.
This resistively produced radiation can be so big that it can explain the
extreme radiation from quasars that are a hundred times more luminous than
ordinary galaxies (e.g.\ \cite{BEL84}).

\begin{figure}[t!]
\includegraphics[height=0.28\textheight]{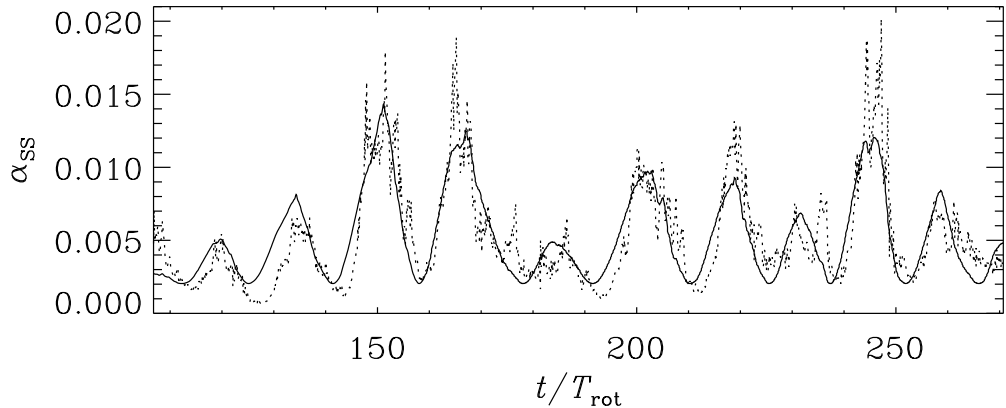}
\caption{Comparison of the evolution of $\alpha_{\rm SS}$ (dotted line)
with a fit \eq{alphafit} to the mean magnetic field (solid line)
using data for Run~A of \cite{BRA96}.}
\label{pfluct}
\end{figure}

The standard theory of disc accretion is quite straightforward provided
the discs can be treated as geometrically thin (e.g.\ \cite{FRA92,CAM97}).
The details of the magnetic stress are assumed not to matter.
Based on dimensional reasoning the sum of the Reynolds and Maxwell
stresses should scale like
\EQ
\left[\overline{u_ru_\phi}-\overline{b_rb_\phi}/(\mu_0\rho_0)\right]
\approx\alpha_{\rm SS}\Sigma\Omega^2H,
\EN
where $\alpha_{\rm SS}$ is the magnetic contribution to the dimensionless
Shakura-Sunyaev parameter \cite{SHA73}, $\Sigma=\int_0^\infty\rho\dd 
z=2\rho_0H$,
$\Omega=\sqrt{GM/R^3}$ is the keplerian velocity at radius $R$, and
$H$ is the scale height.
In practice the Maxwell stress is always larger than the Reynolds stress,
but both tend to add to the stress, so there is no cancelation.
Much of the work on the MRI in discs has focussed on determining the time
averaged value of $\alpha_{\rm SS}$.
In \Fig{pfluct} we show the evolution of $\alpha_{\rm SS}$ for Run~A of
\cite{BRA96}.

Here, much of the time variability comes from the fact that this simulation
develops a large scale field, $\meanBB$, that is oscillatory.
The dependence $\alpha_{\rm SS}(\meanBB)$ can roughly be described via
\EQ
\alpha_{\rm SS}(\meanBB)\approx
\alpha_{\rm SS}^{(0)}+\alpha_{\rm SS}^{(B)}\meanBB^2/B_0^2,
\label{alphafit}
\EN
where $\alpha_{\rm SS}^{(0)}\approx0.002$, $\alpha_{\rm SS}^{(B)}\approx0.06$,
and $B_0^2=\mu_0\bra{\rho}\bra{c_{\rm s}^2}$.
In the original simulations of \cite{BRA95} this
time dependence of the large scale field was associated with an
$\alpha\Omega$ dynamo.
This large scale field also showed migration away from the equatorial plane.
Various approaches have subsequently confirmed the initial result that
$\alpha$ is negative in the northern hemisphere.
This is quite unusual and is probably related to the swirl that comes from the
combined action of magnetic buoyancy and the Coriolis force acting on the
field-aligned flows driven by the $\BB\cdot\nab\BB$ tension force
\cite{BRA98}.

As expected, the dynamo alpha requires the presence of both rotation and
stratification.
The simulations of \cite{HAW96,STO96} without
imposed field did show dynamo action, but they assumed periodicity in the
vertical direction and there was no net stratification and hence no
$\alpha\Omega$ dynamo-type behavior.
Subsequent simulations of \cite{MIL00} had strong stratification
and open boundary conditions.
The vertical extent of the box was also much larger than in the previous
simulation ($|z|\leq5H$).
Nevertheless, no large scale field was observed.
At the moment we can only speculate about the possible origin of this
discrepancy.
One possibility is the fact that they used a numerical resistivity that
allowed very little magnetic helicity to be dissipated.
Subsequent work in the context of helically forced turbulence showed that
a large scale dynamo effect involving the so-called $\alpha^2$ mechanism
(for homogeneous helical turbulence, but no shear) can only saturate on
a resistive time scale \cite{BRA01}.
Whether or not this is indeed the reason for the absence of a large scale
field in the simulations of \cite{MIL00} remains open.

\section{High resolution direct simulations}

Several groups have shown independently that the
MRI leads to sustained fully three-dimensional turbulence
in the presence of an imposed (vertical or toroidal) magnetic field
\cite{HAW95,MAT95}.
Similar simulations have also shown that there is the possibility that
the MRI coupled to the dynamo instability can lead to a doubly-positive
feedback whereby the dynamo produces the magnetic field for the MRI, which
in turn produces the turbulence for the dynamo
\cite{BRA95,HAW96,STO96,ZIE00}.

In all the simulation results published so far, a nonuniform effective
viscosity or some other equivalent numerical procedure has been adopted
in order to maximize the Reynolds number in regions where the flow is
quiescent and to reduce it as much as necessary in regions where the
flow shows strong spatial variations.
These statements also apply to the effective magnetic diffusivity.
Although these procedures are known to provide reasonably safe
approximations to many types of turbulent flows
\cite{HAU04}, there are examples
in the context of helical turbulence where
departures from the ordinary viscosity and magnetic diffusion operators
can lead to major differences compared with the correct solution
\cite{BRA02}.

In accretion discs the main source of heating is viscous and Joule heating.
In the absence of cooling, this would lead to secular heating of the model.
In order to avoid this, it is customary to add a volume cooling
\cite{BRA96}.
In the present model we adopt instead an isothermal gas with constant
sound speed.
We thus solve \Eqss{dlnrhodt}{dudt} using
as initial conditions a simple magnetic field configuration given by
\EQ
A_z=A_0\cos (k_1 x) \cos (k_1 y) \cos (k_1 z),
\EN
where $A_0=0.2$ has been chosen for all our runs.
The initial field strength is chosen rather high so that the wavelength
of the most unstable mode, $v_{\rm A}/\Omega$, is large compared with
the mesh spacing, $\delta x$.
Alternatively, especially for high resolution runs, we use a snapshot
from a lower resolution run and remesh it using interpolation.

Unless noted otherwise, we present the results by measuring time in
units of $\Omega^{-1}$, length in units of $k_1^{-1}$, and density in
units of the initial density, $\rho_0$.
The orbital period is $T_{\rm rot}=2\pi/\Omega$, which is sometimes also
used when the time is given as $t/T_{\rm rot}$.
We have carried out simulations at three different resolutions with three
different viscosities, keeping the magnetic Prandtl number $\Pm=\nu/\eta$
always equal to unity; see \Tab{Tsum}.

\begin{figure}
\includegraphics[height=.3\textheight]{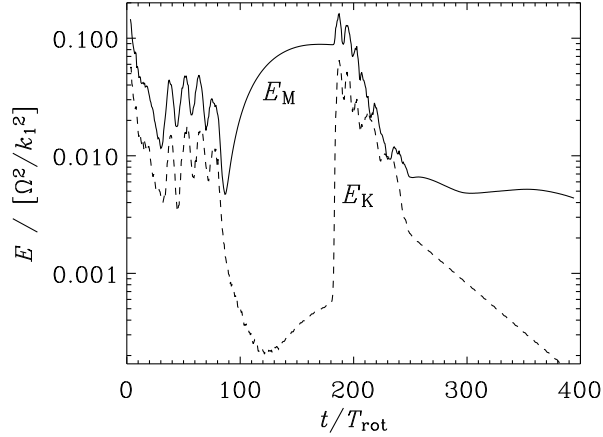}
\caption{
Evolution of kinetic and magnetic energies in a run with
$128^3$ meshpoints. $\nu=\eta=1.5\times10^{-3}\Omega/k_1^2$,
where $k_1=1=2\pi/L$ and $L=2\pi$ is the side length of the simulation
box with uniform aspect ratio.
}\label{pn}\end{figure}

In the run with $128^3$ meshpoints we have been able to run for more
than three hundred orbital times; see \Fig{pn} for a plot showing the
evolution of kinetic and magnetic energies.
After the first 20 orbits the magnetic energy (per unit volume) decays
rapidly and comes then to a halt at around 0.03 (the value quoted in
\Tab{Tsum}).

\begin{table}[t]\caption{
Mean magnetic and kinetic energies
(per unit volume and in units of $E_0\equiv\rho_0\Omega^2/k_1^2$)
for runs at different resolution and different viscosities (for unit
magnetic Prandtl number, $\nu=\eta$ in units of $\Omega/k_1^2$).
The nondimensional stress is here given as $\tilde\alpha_{\rm SS}\equiv
[\bra{u_xu_y}-\bra{b_xb_y}/(\mu_0\rho_0)]k_1^2/\Omega^2$.
For runs with regular viscosity, the magnetic Reynolds number is defined
as $R_{\rm m}=u_{\rm rms}/(\eta k_1)$.
The duration of the run is indicated to give some idea about the
statistical significance of the data.
In all runs the normalized sound speed is $c_{\rm s}k_1/\Omega=5$.
}\vspace{12pt}\centerline{\begin{tabular}{ccccccccc}
Res. & $\nu$ ($=\eta$) & $\bra{E_{\rm M}}/E_0$
& $\bra{E_{\rm M}}/\bra{E_{\rm K}}$ & $R_{\rm m}$
& $\tilde\alpha_{\rm SS}$ & $t_{\rm run}/T_{\rm rot}$ & status \\
\hline
 16 hyp & $7.5\times10^{-5}$ & 0.073 & 3.5 & --  &   --  &1027 & decay\\ 
 32 hyp & $3.5\times10^{-6}$ & 0.111 & 2.1 & --  & 0.056 & 180 & turb \\ 
128 dir &  $15\times10^{-4}$ & 0.031 & 3.0 &  90 &   --  & 450 & decay\\ 
256 dir & $5.0\times10^{-4}$ & 0.013 & 2.5 & 200 & 0.005 & 117 & turb \\ 
512 dir & $2.5\times10^{-4}$ & 0.019 & 2.5 & 500 & 0.010 &   6 & turb \\ 
\label{Tsum}\end{tabular}}\end{table}

\begin{figure}[t]
\includegraphics[height=.3\textheight]{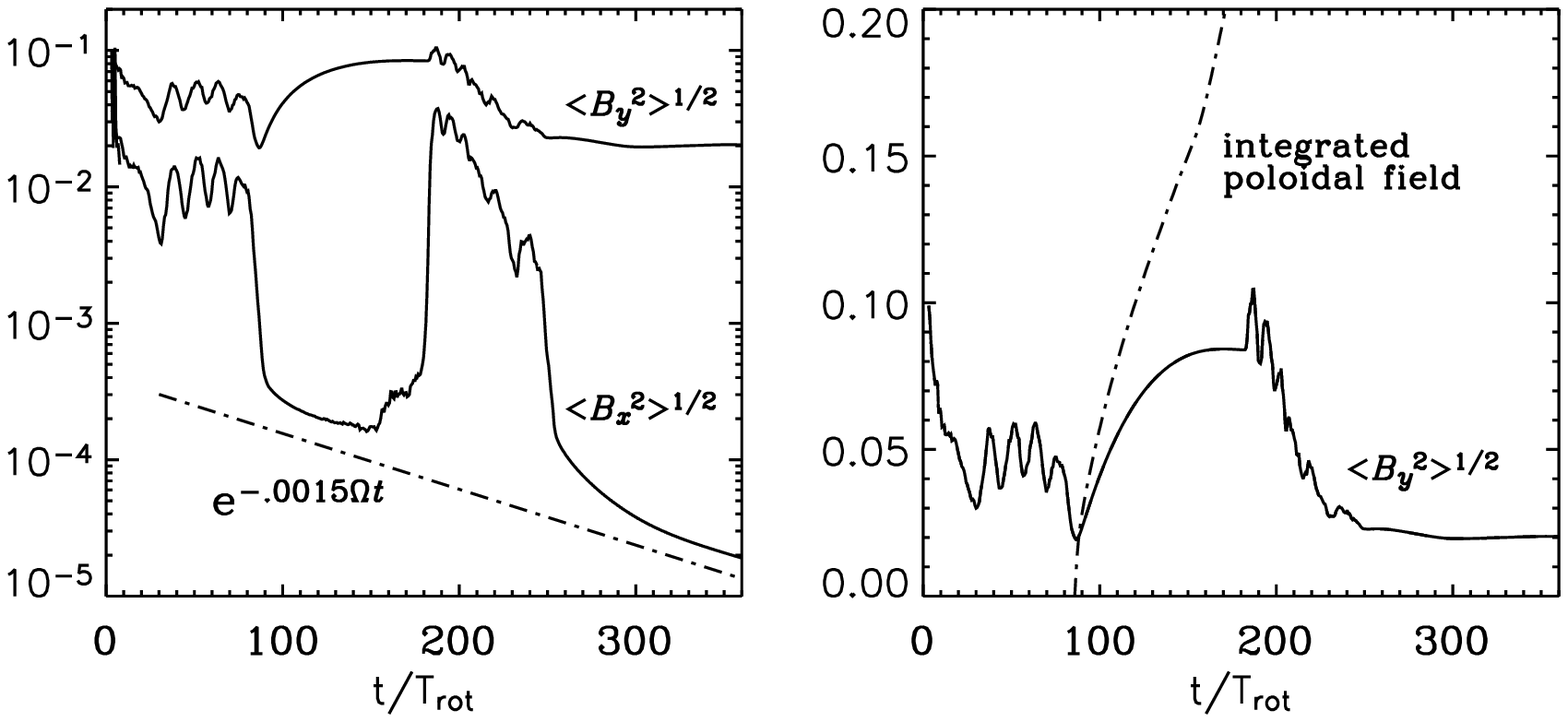}
\caption{
Comparison of $\bra{B_x^2}^{1/2}$ with $\bra{B_y^2}^{1/2}$ (left panel)
and $\int{3\over2}\Omega\bra{B_x^2}^{1/2}\dd t$ (shown as dash-dotted line)
with $\bra{B_y^2}^{1/2}$ (right panel).
$128^3$ meshpoints. $\nu=\eta=1.5\times10^{-3}\Omega/k_1^2$.
}\label{pgrowth}\end{figure} 

Both kinetic and magnetic energies vary approximately periodically in
intensity, but then at $t\approx100T_{\rm rot}$
the kinetic energy of the turbulence drops drastically by two
orders of magnitude.
The magnetic field also drops at first, but starts then to grow
approximately quadratically in time, corresponding to a linear increase
in the rms field strength.
This increase is readily explained by the presence of a residual
cross-stream rms magnetic field.
Since in the simulation the magnetic vector potential is periodic (or
shearing-periodic in the $x$ direction), there is no net flux, so the
most slowly decaying mode has the wavenumber $k/k_1=1$.
This mode decays at a rate $\eta k_1^2\approx0.0015$.
This is indeed consistent with the data (see \Fig{pgrowth}).
In \Fig{pgrowth} we also demonstrate that the time integrated
cross-stream rms magnetic field provides a rough estimate of the
resulting toroidal magnetic field by winding up the poloidal field.
Eventually, the field strength reaches an amplitude that is large
enough for the MRI to set in ($t\approx200T_{\rm rot}$).
This leads to renewed turbulence for some 50 orbits, but then the
magnetic and kinetic energies have dropped so much that the MRI
becomes unable to sustain the turbulence
(see \Fig{pn} at $t\approx200T_{\rm rot}$).

To make contact with earlier work we now use hyperviscosity, i.e.\ the
differential operator $\nabla^2$ is replaced by $\nabla^6$.
In numerical turbulence, hyperviscosity has previously been found to
give rise to an artificially strong bottleneck effect, i.e.\ a shallower
spectrum than $k^{-5/3}$ near the dissipative subrange.
However, more recent work has shown that the width of the bottleneck is
always around one order of magnitude both for direct and hyperviscous
simulations, and that the actual inertial range is not affected by the
presence of hyperviscosity \cite{HAU04}.

\begin{figure}[t]
\includegraphics[height=.3\textheight]{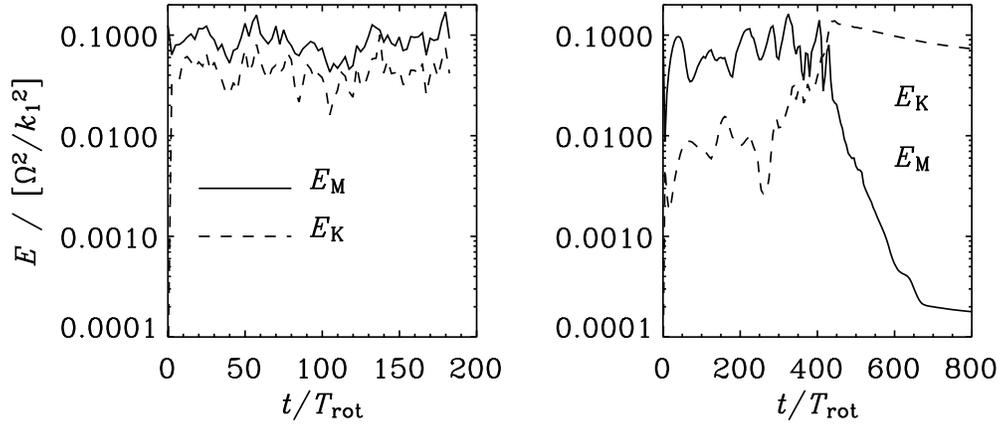}
\caption{
Evolution of kinetic and magnetic energies in runs using
hyperviscosity and hyperresistivity with
$32^3$ meshpoints, $\nu_3=\eta_3=3.5\times10^{-6}\Omega/k_1^6$ (left), and
$16^3$ meshpoints, $\nu_3=\eta_3=7.5\times10^{-5}\Omega/k_1^6$ (right).
}\label{pn16+32}\end{figure}

Using this as general reassurance of the usefulness of turbulence
simulations with hyperviscosity, we now proceed using this technique
in the present case of shearing sheet accretion disc turbulence.
In the following, however, we shall only consider a rather small number
of meshpoints, in which case there is no inertial range anyway.
Therefore, such hyperviscous simulations cannot be considered as an
approximation to the full equations, but they should really only
be regarded as an illustrative model resembling features of a high
resolution model.
At a resolution of just $32^3$ we have in this way been able to run
for 300 hundred orbits and found a steady level of turbulence
(left hand panel of \Fig{pn16+32}).
The absence of dramatic variations of the turbulence intensity suggests
that such variations were an artifact of still too low resolution in the
direct simulations presented above.
This is supported by another lower resolution run (only $16^3$ meshpoints)
with hyperviscosity and hyperdiffusivity shown in the right hand panel
of \Fig{pn16+32}.
We see that the behavior is similar to what we see in \Fig{pn}, suggesting
that the reason for the disappearance of the MRI is really just that the
Reynolds number dropped below a certain critical value.

In the rest of this section we consider the properties of the energy
spectra for the direct simulations with high resolution of $256^3$
and $512^3$ meshpoints.
It turns out that the magnetic energy spectrum increases with wavenumber
like $k^{2/3}$.
The kinetic energy scales approximately like $k^{-1/3}$.
For neither of the two spectra do we have a reasonable explanation.
Obviously, the larger the resolution, the harder it becomes to run for
a long time.
Our run with $256^3$ meshpoints has run for nearly 120 orbits, but this
may not be enough to exclude conclusively a subsequent decay.

\begin{figure}[t]
\includegraphics[height=.3\textheight]{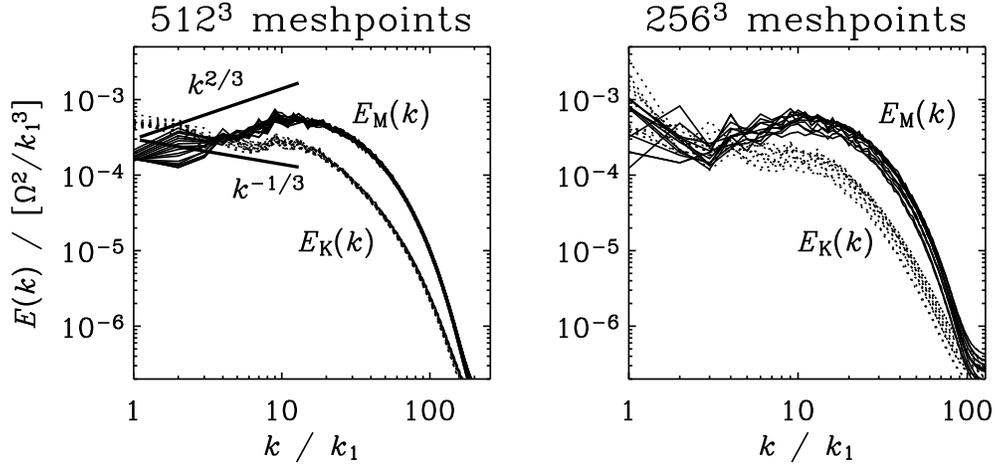}
\caption{
Kinetic and magnetic energy spectra for the run with
$512^3$ meshpoints, $\nu=\eta=2.5\times10^{-4}\Omega/k_1^2$ (left) and
$256^3$ meshpoints, $\nu=\eta=5\times10^{-4}\Omega/k_1^2$ (right).
}\label{pspecall}\end{figure}

\begin{figure}[t]
\includegraphics[height=.3\textheight]{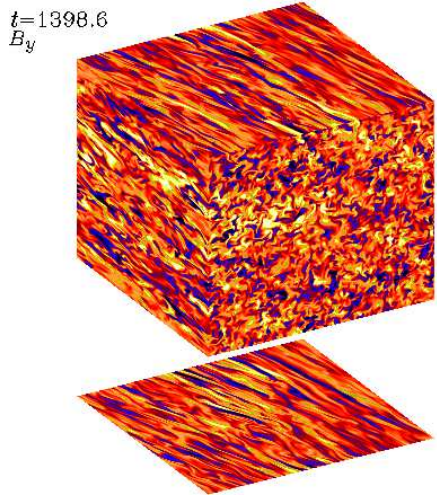}
\caption{
Toroidal magnetic field component displayed on the
periphery of the computational domain (color coded).
$512^3$ meshpoints. $\nu=\eta=2.5\times10^{-4}\Omega/k_1^2$.
}\label{BH512By}\end{figure}

The simulations presented above may help to determine a critical value
of the magnetic Reynolds number, $R_{\rm m}^{\rm(crit)}$, above which
dynamo-generated turbulence becomes possible.
If the turbulence in the simulation with $256^3$ meshpoints remains indeed
sustained, we expect $R_{\rm m}^{\rm(crit)}$ to be between 100 and 200.
However, the values of $\alpha_{\rm SS}$ are not converged in any of
the simulations in that its value increases with resolution, and it
is also much larger in the simulations with hyperviscosity than with
ordinary viscosity.

\section{Spherical Couette flow (preliminary)}

\begin{figure}[t]
\includegraphics[height=.3\textheight]{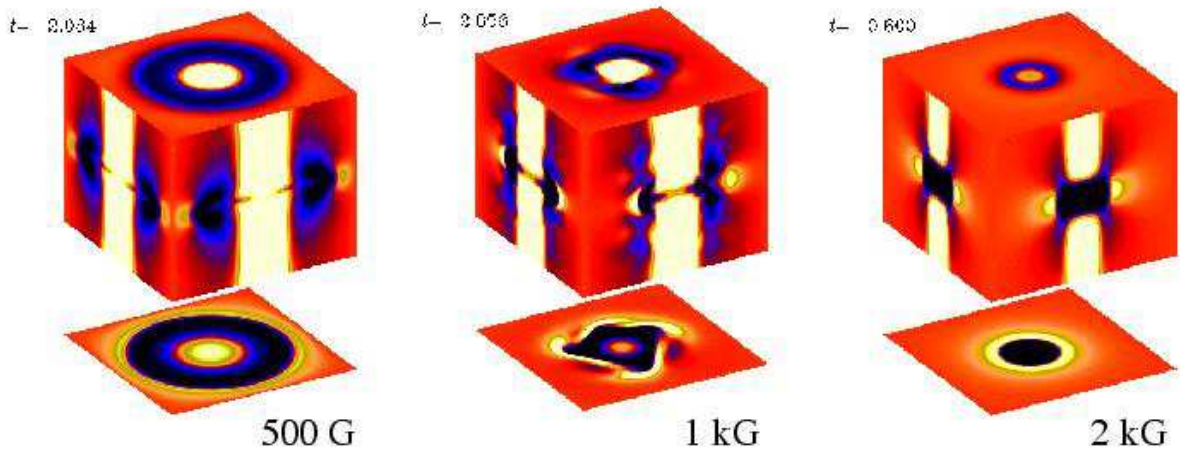}
\caption{
Images of the vertical component of the magnetic field, $B_z$, for three
different values of the imposed field, $B_0$.
The images shown on the two side walls of the box go actually through
the axis ($x=y=0$), the top plane goes through $z=0.09\m$ and the bottom
plane goes through $z=0$.
}\label{images}\end{figure}

In order to better understand the experimental verification of
the MRI in spherical Taylor-Couette flow by the Maryland group
\cite{SIS04}, it is desirable to perform
simulations that are related to the setup used in the laboratory.
We know already that cylindrical Taylor-Couette flow can well be
modeled by embedding the flow in a cartesian box \cite{DOB02}.
We also know that simulations of the geodynamo (convection-driven
dynamos in a spherical shell) can successfully be modeled by embedding
the spherical shell in a box \cite{MCM04}.
It is therefore natural to attempt modeling of the Maryland MRI
experiment using the same setup.
In \Fig{images} we present some preliminary results of spherical Couette
flow embedded in a cartesian box with an imposed axial magnetic field
of given strength.
We work in SI units and chose parameters that are closely related to the
experimental setup.
A sphere of radius $R=0.15\m$ is embedded in a box of size
$(2\times0.18\m)^3$.
The sphere spins with an angular frequency of $\Omega=300\s^{-1}$.
Outside the sphere the velocity is damped to zero at rate
$\tau_{\rm damp}^{-1}=200\s^{-1}$.
The density is initially uniform and equal to $\rho=930\kg\m^{-3}$.
For reasons of computational simplicity
our simulation is weakly compressible, and we use an isothermal equation
of state (ratio of specific heats is $\gamma=1$) with sound speed
$c_{\rm s}=50\m\s^{-1}$.
For the present simulations no attempt is made to use realistic values
of the kinematic viscosity and the magnetic diffusivity; both are chosen
equally big with $\nu=\eta=2\times10^{-3}\m^2\s^{-1}$.
The field strength is varied between $500\G$ and $2\kG$.

It turns out that for weak and strong fields the magnetic field is
completely axisymmetric.
For an intermediate field strength of $B_0=1\kG$ both the flow and the
magnetic field become nonaxisymmetric.

As anticipated by R.\ Hollerbach (private communication), the main reason
this simulation produces nonaxisymmetric flows is the modification of the
background flow in such a way that it becomes Rayleigh unstable, i.e.\
$q\equiv-\dd\ln\Omega/\dd\ln r>2$.
On the other hand, this does not seem to be the case in the Maryland
experiment where the flow remains close to keplerian, i.e.\ $q\approx1.5$.

\section{Conclusions}

The main outcome of the magnetorotational instability is by now well
known and its importance for astrophysical discs is obvious.
Many details, however, remain far from clear due to the fact that
many of the numerical shearing sheet simulations invoke artificial
viscosity which is hard to quantify.
Thus, we have no clear idea about the critical value of the magnetic
Reynolds number required for the combined magnetorotational and dynamo
instabilities.
The present work suggests that the number is around 200, and certainly
above 100, but obviously a more quantitative study would be desirable.
Furthermore, details of the nonlinear mode interactions are not well
understood.
This would be very useful for obtaining a clearer picture of how in the
shearing sheet approximation the nonaxisymmetric MRI leads to sustained
growth.
The {\sc Pencil Code} is well suited for such studies, and it is publicly
available, so it may be hoped that the remaining gaps in our understanding
will soon be filled by new members of the scientific community.

Regarding spherical Couette flow, the present approach using embedded
box simulations may not be optimal, and dedicated codes for spherical
geometry may be advantageous.
However, any independent verification using different methods is still
often useful.
Returning to the astrophysical context, it is important to assess the
relative importance of dynamo-generated fields relative to those generated
on a more global scale in the rest of the disc.
This question can only be appropriately in a fully global simulation.
Indeed, significant progress has recently been made in this direction
\cite{HAW00,HAW01,HAWK01,DEV03}.

\begin{theacknowledgments}
We thank the Danish Center
for Scientific Computing for granting time on the Horseshoe cluster,
and the Norwegian High Performance Computing Consortium (NOTUR)
for granting time on the parallel computers in 
Trondheim (Gridur/Embla) and Bergen (Fire).
This work has been carried out in part at the Isaac Newton Institute
in Cambridge.
\end{theacknowledgments}


\bibliographystyle{aipproc}   

\bibliography{biblio}

\end{document}